\documentclass[pre,twocolumn,showpacs,amsmath,amssymb]{revtex4}

\newcommand{\bd}[1]{\mbox{\boldmath$#1$}}
\usepackage{graphicx}

\begin{document}

\title{Rotating Spiral Waves with Phase-Randomized Core in Non-locally Coupled 
Oscillators}

\author{Shin-ichiro Shima}
 \email{s_shima@ton.scphys.kyoto-u.ac.jp}
\author{Yoshiki Kuramoto}
\affiliation{
Department of Physics, Graduate School of Sciences, Kyoto University,
Kyoto 606-8502, JAPAN
}

\date{\today}

\begin{abstract}
Rotating spiral waves with a central core 
composed of phase-randomized oscillators can arise in 
reaction-diffusion systems if some of the chemical components involved are
diffusion-free. This peculiar phenomenon 
is demonstrated for a paradigmatic three-component
reaction-diffusion model. 
The origin of this anomalous spiral dynamics is the 
effective non-locality 
in coupling, whose effect is stronger for weaker coupling. 
There exists a critical coupling strength which is  
estimated from a simple argument. Detailed mathematical and numerical 
analyses are carried out in the extreme case of weak coupling
for which the phase reduction 
method is applicable. Under the assumption that the mean field pattern
keeps to rotate steadily as a result of a
statistical cancellation of the incoherence,
we derive a functional
self-consistency equation to be satisfied by 
this space-time dependent quantity.
Its solution and the resulting effective frequencies of the 
individual oscillators
are found to agree excellently with the numerical simulation.
\end{abstract}


\maketitle

\section{INTRODUCTION}
Rotating spiral waves represent a most universal form of patterns
appearing in reaction-diffusion systems and other dissipative
media of oscillatory and excitable nature. Most of the recent
experimental and theoretical studies on rotating spiral waves 
in reaction-diffusion systems have
focused on their complex behavior such as core meandering
in 2D\cite{21,22}, manipulation of the pattern using photo-sensitive BZ
reaction\cite{23,24}, and the topology and dynamics of the singular
filaments of 3D scroll waves\cite{15,16,17}. Slightly deviated from this mainstream,
the possibility of a new type of spiral dynamics 
caused by a universal mechanism 
was proposed recently by the present authors\cite{1}. 
This is characterized by the appearance of a local group of oscillators
near the center of spiral rotation 
where the oscillators behave individually rather than collectively.
A simple class of three-component reaction-diffusion 
systems involving two diffusion-free components and an extra diffusive
component proved to exhibit
this type of anomaly. 
Here the last component plays the role of a 
coupling
agent allowing the otherwise independent local oscillators to communicate
with each other, where the communication takes place 
non-locally.
The crucial parameter to this peculiar spiral dynamics is the strength
of the non-local coupling.
If it is
sufficiently large, the characteristic wavelength of the pattern,
especially the radius of the spiral core, 
becomes
longer than the coupling radius. Consequently, the coupling becomes 
effectively 
local, i.e., diffusive, and there is nothing peculiar about the
resulting spiral pattern.
As the coupling
becomes weaker, in contrast, its non-local nature becomes stronger, and
finally a small group of phase-randomized oscillators starts to be created 
near the center of rotation.
We find in the present paper that under certain conditions the
phase-randomized core is stationary in a statistical sense.
This allows us to formulate a statistical theory with which
the entire system dynamics, collective and individual, can completely be
specified. Actually, the exact self-consistent  
theory developed here provides a rare example of statistical theories
associated with large systems of limit-cycle oscillators when spatial
degrees of freedom are involved. 

The organization of the present paper is the following. 
In Sec.~II, we start with a brief review of some general features of the 
three-component reaction-diffusion model introduced earlier, 
and show how it is reduced
to a two-component system of non-locally coupled oscillators. 
Then, adopting a specific model for the local
oscillators, we present some results of our numerical simulation
revealing the fact that the spiral core can be coherent or incoherent
depending on the coupling parameter. We shall also see that the critical
coupling strength associated with the onset of incoherence
can be estimated from a simple argument.
Section III and IV, each devoted to numerical and mathematical analyses,
are concerned with the special situation
where the coupling is sufficiently weak. 
Then the so-called phase reduction method is
applicable, by which a phase oscillator model with non-local coupling
is derived. What is remarkable is the fact that, 
as opposed to the conventional view, description of the spiral dynamics
in terms of the phase oscillator model does not
lead to a topological contradiction but can even provide 
its precise description.
Using this non-local phase model, we develop a mean field theory
similar to Kuramoto's 1975 theory on the onset of collective synchronization
in globally coupled oscillators\cite{5}. The present mean field theory 
is based on the assumption of 
steady rotation of the mean field pattern. Owing to this assumption, 
we can derive 
a functional 
self-consistency equation to be satisfied by the mean field.
Numerical solution of this functional
equation is confirmed to agree exceedingly well with the simulation results.
\section{REACTION-DIFFUSION MODEL AND ITS REDUCED FORM}
\subsection{Effective Non-locality in Reaction-Diffusion Systems}
Consider a
three-component reaction-diffusion system of the following form.
\begin{align}
\partial_tX(\bd r,t)&=f(X,Y)+K(B-X), \label{1}\\ 
\partial_tY(\bd r,t)&=g(X,Y), \label{2}\\ 
\tau\partial_tB(\bd r,t)&=-B+D\nabla^2 B+X.\label{15}
\end{align}
The system is supposed to extend sufficiently in two dimensions.
The above model has recently been used as a paradigmatic model 
for the study of 
various aspects of self-oscillatory fields where the effective non-locality
in coupling plays a crucial role\cite{1,6,8}. 
The first two equations with $K=0$
represent a local limit-cycle oscillator. 
Our system may therefore be
interpreted as a continuum limit of a large assembly of oscillators
without direct mutual coupling which are suspended in a diffusive chemical with concentration $B$. The last quantity plays the role
of a coupling
agent only by which the local oscillators can mutually communicate. 
For simplicity, a cross coupling
between the local oscillators and the diffusive field has been introduced 
in a linear form and only between $X$ and $B$.
The coupling term $K(B-X)$ 
may equivalently be replaced 
with a more natural form 
$KB$ if $f(X,Y)$ is suitably redefined,
but we will work with the first form for its mathematical convenience 
to be seen later.

Our system, possibly with various modifications and generalizations,
bears some
resemblance to biological populations of oscillatory and excitable
cells such as suspensions of yeast cells under glycolysis and slime mold
amoebae in a certain phase of their life cycle\cite{12,13,14}. 
One may also note some similarity of the above model
to the recently
developed version of the Belousov-Zhabotinsky reaction
using water-in-oil AOT micro-emulsion\cite{18,19,20}. Some of the 
interesting theoretical aspects of our reaction-diffusion model 
have already been reported
\cite{1,6,8}.

When the characteristic time scale of $B$, denoted by $\tau$,
is sufficiently small,
this component can be eliminated
adiabatically by solving the equation
\begin{equation}
\label{3b}
0=-B+D\nabla^{2}B+X.
\end{equation}
The solution of the above equation 
is expressed in terms of the Green function $G(\bd{r})$
in the form 
\begin{equation}
\label{3}
B(\bd r,t)=\int G(\bd{r}-\bd{r}')X(\bd{r}',t)d\bd{r}'. 
\end{equation}
If our system is infinitely
extended, $G(\bd{r})$ is radially symmetric, and for spatial dimension
two it is given by a modified 
Bessel function 
of the second kind with the characteristic length scale $r_0=\sqrt{D}$, i.e.,
\begin{equation}
\label{9}
G(r)=\frac{1}{2\pi r_0^2}\;K_{0}\!\!\left(\frac{r}{r_{0}}\right),\quad r=|\bd{r}|.
\end{equation}
Note that the above $G(r)$ satisfies 
the normalization condition $\int G(r)d\bd{r}=1$, and behaves asymptotically
as $G(r)\sim\exp(-r/r_{0})/\sqrt{r/r_{0}}$ for $r\gg r_{0}$. 
We may call $B(\bd r,t)$ the space-time dependent {\em mean field}
because this quantity roughly represents a
mean value of $X(\bd{r}',t)$ over a circular domain with the radius
of $O(\sqrt{D})$ centered at $\bd{r}$.
Our original reaction-diffusion system has now
been reduced to the system of Eqs.~\eqref{1}, \eqref{2}
and \eqref{3}, which represents a two-component oscillatory field with
non-local coupling.  

Suppose that we change parameter $K$, which measures, in terms of
the reduced system, the strength of the
non-local coupling.
If $K$ is sufficiently large, the characteristic wavelength of the pattern,
denoted by $l_p$,
will be far longer than the coupling radius (as is justified below).
Then the long-wavelength approximation can applied to Eq.~\eqref{3},
giving $K(B-X)\simeq\tilde{D}\nabla^2 X$,
where $\tilde{D}=KD$. Thus, the non-local coupling practically reduces to
a diffusive coupling in this strong-coupling case.
One may check the consistency of the above
argument by noticing the fact that
 the result of this diffusion-coupling approximation itself 
tells that
$l_p$ estimated from Eq.~\eqref{1} scales like $l_p\sim \sqrt{\tilde{D}}=\sqrt{KD}$. Thus, 
sufficiently large $K$
implies $l_p$
larger than the coupling radius $\sqrt D$, so that our long-wavelength 
assumption proves to be 
consistent. In any case, our system for strong coupling reduces 
to a
two-component reaction-diffusion
system, which, however, is of our little concern in the present paper.
 
The situation of our interest is the opposite case in which
$K$ is so small that  $l_p$ 
becomes comparable with or even smaller than the
coupling radius of $O(\sqrt D)$. Then the diffusion-coupling approximation 
breaks down, 
and the system comes to behave in an unusual manner.
It should be noted here that the evolution equations themselves
are free of characteristic length scale below the coupling radius. Therefore,
once $l_p$ comes to fall within the coupling radius, or equivalently,
once spatial variations with wavelengths smaller than the coupling radius 
are generated spontaneously, then there is no reason why 
spatial variations of even smaller 
wavelengths
should not occur. We suspect therefore 
that the kind of anomaly of our concern might be  
characterized
by a fragmentation of the pattern down to infinitesimal spatial scales.
This is actually the case, which we show below by presenting
some numerical results on 
the non-locally coupled system given by 
Eqs.~\eqref{1}, \eqref{2} and \eqref{3} 
with specific forms for $f$ and $g$. 
\subsection{Case of the FitzHugh-Nagumo Oscillators}
As a simple model for the local oscillators, let us consider the
FitzHugh-Nagumo model given by
\begin{equation}
f=\sigma^{-1}\{(X-X^3)-Y\}, \quad g=aX+b.
\label{4}
\end{equation} 
We fix the parameter values as $a=1.0$, $b=0.2$ and $\sigma=0.1$, 
so that
the system is well in the self-oscillatory regime.

We carried out a numerical analysis on a non-locally coupled
field of oscillators described by
Eqs.~\eqref{1}, \eqref{2}, \eqref{3} and
\eqref{4}.
The system is defined over the square domain $x,y\in [0,L]$ where $B$ satisfies
the free boundary conditions,
namely, the vertical component of $\nabla B$ to the boundaries vanishes.
Thus, the Green function $G$ differs in this case
from the form given by Eq.~\eqref{9}
especially near the boundaries. In practical numerical simulation, 
our continuous space 
was replaced with a square lattice of oscillators of $N\times N$ 
lattice points, a typical value of $N$ being 2048. At each time step,
$B$ was calculated from
Eq.~\eqref{3}, or equivalently Eq.~\eqref{3b}, by means of a spatial
Fourier transform. The 4th order Runge-Kutta scheme was adopted
for the time-integration of Eqs.~\eqref{1} and \eqref{2}.

\begin{figure}
\includegraphics[width=8.6cm]{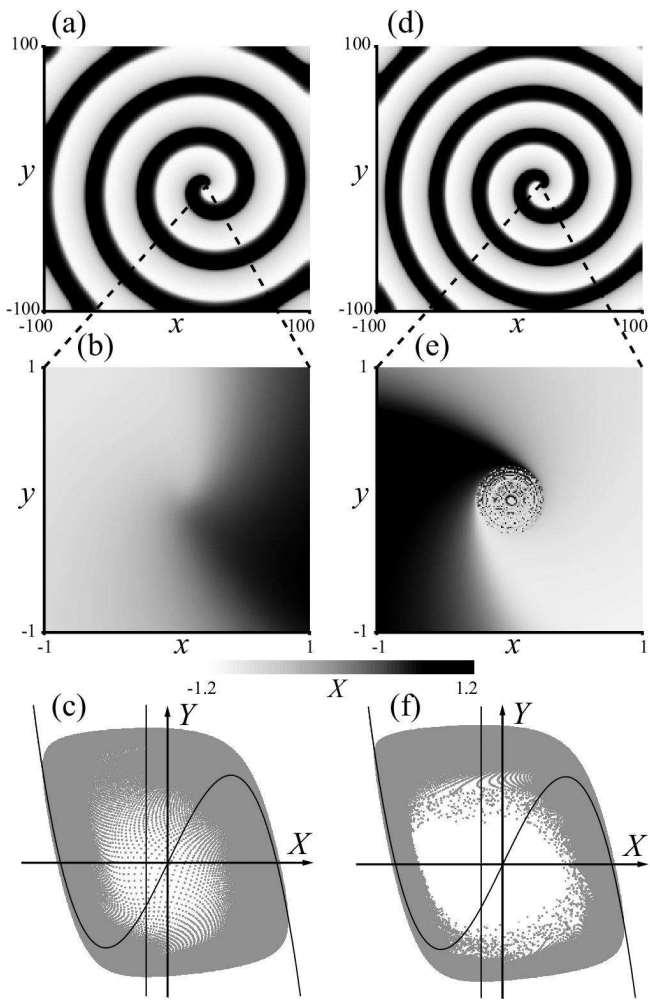}
\caption{\label{fig1} 
Spiral patterns exhibited by non-locally coupled FitzHugh-Nagumo oscillators
for two representative cases of 
strong coupling ((a),(b) and (c), $K=10.0$) and 
weak coupling 
((d),(e) and (f), $K=5.0$). Other parameters are fixed as
$a=1.0$, $b=0.2$, $\sigma=0.1$ and $D=1$.
(a) and (d): Overall patterns of the component $X$ displayed in gray scale.
(b) and (e): Their structures near the core.
(c) and (f): Corresponding phase portraits in the $X$-$Y$ plane, where 
the nullclines $f(X,Y)=0$ and $g(X,Y)=0$ are also indicated. }
\end{figure}

Some numerical results for two representative values of $K$ are illustrated
in FIG.~\ref{fig1}.
Figure~\ref{fig1} (a) to (c) correspond to the case of large $K$. 
They respectively show an overall spiral pattern, its blowup near the center
of rotation, and the phase portrait of the pattern in the $X$-$Y$ plane.
The last quantity is given by a set of $N^2$ points in the 
$X$-$Y$ plane each representing the state of a local 
oscillator at a given time. In usual reaction-diffusion systems such as
those modeled with two-component reaction-diffusion equations,
the phase portrait associated with a spiral pattern 
is considered to form a simply connected object
involving a phase singularity. This is a natural consequence of the
homeomorphism which is supposed to characterize
the mapping between the physical space and the state space.
The same property seems to hold in the 
present case of large
$K$, and this is consistent with the fact already noted 
that for sufficiently strong coupling our system reduces to a
two-component reaction-diffusion system.

Figure~\ref{fig1} (d) to (f) correspond to the case of small $K$.  The overall
spiral pattern does not seem qualitatively 
different from that for large $K$. As is clear from Fig.~\ref{fig1} (e),
however, closer observation of the
core structure reveals a completely
new feature of the pattern. This is the appearance of a group of
oscillators near the center of rotation where the oscillators seem to behave 
individually rather than collectively. The corresponding phase portrait,
which is 
presented in FIG.~\ref{fig1} (f), no longer seems to tend to a simply 
connected object
in the continuum limit. The hole created in the phase portrait gives a
clear indication of the 
breakdown of the homeomorphism mentioned above. It may alternatively be said that  
a pair of local oscillators situated 
infinitely 
close to each other
are not always so close in the state space, which says nothing but
a loss of spatial continuity of the pattern. At the same time,
the phase singularity, 
which is generally considered as a central characteristic
shared by spiral patterns, seems to be lost, i.e., the pattern
no longer seems to involve a 
special local oscillator for which the phase cannot
be defined.

The origin of the spiral core anomaly of this kind may 
qualitatively be understood
in the following way.
Our primary question is why the core region is the 
most fragile part of the pattern with respect to the collapse of
spatial continuity.  
In order to see why, it is convenient to look upon Eqs.~\eqref{1} and \eqref{2}
as describing a single oscillator driven by a forcing field $B$ whose spatial
variation is expected to be relatively smooth from its definition given by
Eq.~\eqref{3b}.
Wherever the oscillation amplitude of $B$ is sufficiently large,
the oscillators  will individually synchronize with 
the motion of $B$, so that a local group of such oscillators will
mutually synchronize also. The corresponding local 
pattern will then look continuous and 
smooth. This is considered to be the case for those oscillators 
far apart from the central core, because 
the oscillation amplitude of $B$ there should be relatively large. 
In contrast, close to the central
part of the pattern, where the oscillation amplitude 
of $B$ should be relatively small, synchronization becomes more
difficult. Loss of mutual synchrony implies the appearance of a group of
phase-randomized oscillators.

\subsection{Estimation of $K_c$}
From the numerical data presented in FIG.~\ref{fig1}, 
one may expect the existence of a critical
value of the coupling strength, denoted as $K_c$, associated with 
the onset of incoherence.
We now try to estimate $K_c$ for our system of 
non-locally coupled FitzHugh-Nagumo 
oscillators given
by Eqs.~\eqref{1}, \eqref{2}, \eqref{3} and \eqref{4},
where the spatial extension is supposed to be infinite. 
Consider first the situation where 
the coupling is large enough for the system to 
sustain a rigidly rotating spiral wave with 
sufficient spatial smoothness. The corresponding solution is
represented by 
$\bd A_s(\bd r,t)=(X_s(\bd
r,t),Y_s(\bd r,t))$. Let the center of rotation be at $\bd r=0$.
By assumption, the oscillator right at the center is motionless, i.e.,
$\bd A_s(0,t)=(X_c,Y_c)$, where $X_c$ and $Y_c$ are time-independent. 
Our question is at which value of $K$ this fixed point becomes
unstable and the oscillator there 
starts to oscillate.
To consider this problem, it is convenient to work with the
aforementioned mean-field 
picture by which we look upon 
the local oscillators as being subject to a common space-time 
dependent field $B$. The mean field pattern should also rotate rigidly around
$\bd r=0$, so that the central oscillator is subject to a constant forcing
$B(0)$. The system of Eqs.~\eqref{1} and \eqref{2} describing this particular
oscillator form an autonomous two-dimensional dynamical system, so that once
$B(0)$ is known the value of the fixed point $(X_c,Y_c)$ and its stability
will easily be found.
The value of $B(0)$ can actually be estimated from Eq.~\eqref{3}
by developing $X_s(\bd{r},t)$ into a Taylor series about $\bd r=0$, which
is allowed owing to the assumed smoothness of the pattern.
It is clear that, as a result of the isotropy of the coupling function $G$, 
there is no contribution to $B(0)$ from the first order expansion terms.
If the contribution from the second order terms is negligible, i.e., if the
nonlinear variation of $X_s$ within the coupling range about $\bd r=0$
is negligible, then we may simply put $B(0)=X_c$. With this approximation, 
it is clear from
Eqs.~\eqref{1} and \eqref{2} that the fixed point $(X_c,Y_c)$ is identical with
the intersection of the nullclines $f=g=0$, i.e., the unstable fixed point
of the local oscillators. 
Its linear stability is also easy to analyze.
The result is that the critical coupling strength is given by 
$K_c=(1-3X_{c}^2)/\sigma$ below which the fixed point $(X_c,Y_c)$ 
becomes Hopf unstable.
Applying the values of $a$, $b$ and $\sigma$ used in our numerical 
simulations,
we obtain $K_c=8.8$. This value of $K_{c}$
is consistent with our direct numerical
simulation, although its precise numerical determination
is yet unavailable.

\begin{figure}
\includegraphics[width=8.6cm]{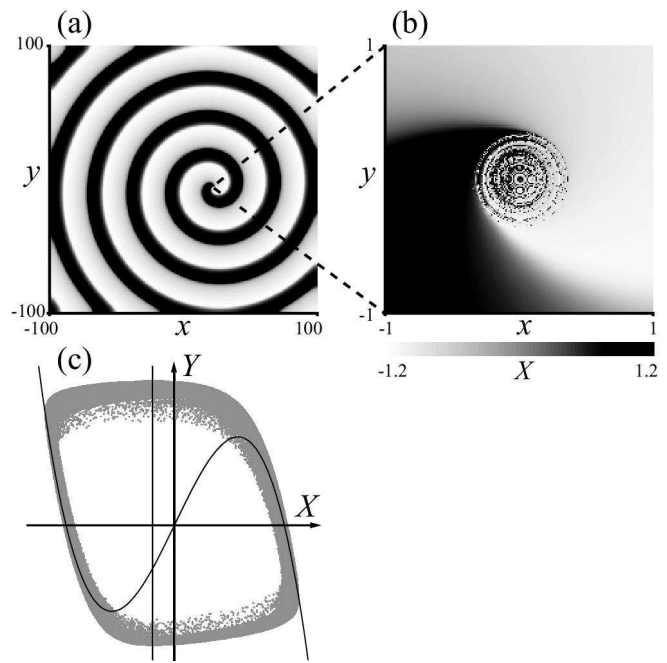}
\caption{\label{fig2} 
Spiral patterns exhibited by non-locally coupled FitzHugh-Nagumo oscillators
with $K=2.0$, presented in a similar manner to FIG.~\ref{fig1}. 
For this value of $K$, the amplitude degrees of freedom
become almost irrelevant.}
\end{figure}

\section{SPIRAL DYNAMICS IN NON-LOCALLY COUPLED PHASE OSCILLATORS}
In order to look into the nature and origin of our anomalous spiral dynamics
in further
detail, we now consider the situation where the coupling 
is much weaker than $K_c$.  Figure~\ref{fig2} shows some results from 
our numerical simulation
carried out for $K=2.0$, presented in a similar manner to FIG.~\ref{fig1}.  
While there seems nothing
unusual about the overall spiral pattern, the corresponding phase portrait
forms a ring with a relatively thin periphery, which is 
totally unlike a simply connected object. 
We may alternatively say 
that the oscillator amplitude everywhere takes almost a full value.
This is exactly the situation where the so-called phase description
is applicable. In fact, as we see later in this section,
a simple phase
oscillator model with non-local coupling can develop a spiral pattern
with phase-randomized core similar to the above. 

We now present a brief review of the phase reduction method\cite{9} in the form
appropriate for the present purposes.
Each of our local oscillators without coupling
is described by a two-dimensional
dynamical system $d\bd{A}/dt=\bd{F}(\bd{A})$, where $\bd{A}=(X,Y)$ and
$\bd{F}=(f,g)$. Let its stable time-periodic solution with frequency $\omega$
be given  by $\bd{A}_0(\omega t)=(X_0(\omega t),Y_0(\omega t))$, which is
a $2\pi$-periodic function of $\omega t$. The corresponding limit-cycle orbit
is represented by $\cal C$.
Phase $\phi$ associated with this oscillator
must be defined outside $\cal C$ as well as on $\cal C$.
Most conveniently, it is defined in such a way that
the free motion of the oscillator satisfies
$d\phi/dt=\omega$ regardless of initial conditions.  
This requires that $\phi$ as a scalar field $\phi(\bd{A})$ satisfies the identity
$\mbox{grad}_{\boldsymbol{A}}\phi\cdot\bd{F}(\bd{A})=\omega$.
The whole $X$-$Y$ plane is then filled with equi-phase lines which
are called isochrons, one of which is chosen to correspond to
the zero phase. 
Corresponding to each  $\phi$-value, a point $\bd{A}_{0}(\phi)$
on $\cal C$ is determined uniquely, which says nothing but the fact that
an isochron and $\cal C$ intersect at a
single point.
 
\begin{figure}
\includegraphics[width=8.6cm]{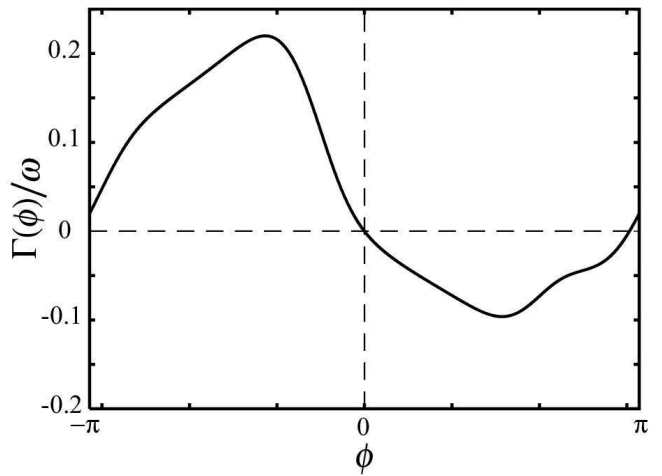}
\caption{\label{fig3} 
Phase-coupling function $\Gamma(\phi)$ versus $\phi$ 
for coupled FitzHugh-Nagumo 
oscillators. 
This quantity can be used for the study of non-locally coupled phase 
oscillators given by Eq.~\eqref{7}.}
\end{figure}

When the non-local coupling is introduced, the equation for each local
oscillator is modified as
\begin{equation}
\label{5}
\partial_t\bd A(\bd r,t)=\bd F(\bd A)+
\bd{p}(\bd{r},t),
\end{equation}
where
\begin{eqnarray*}
&&\bd{p}(\bd{r},t)=(p_{X}(\bd{r},t),0), \nonumber \\
 &&p_{X}(\bd{r},t)=K\int G(\bd{r}-\bd{r}')\left[X(\bd{r}',t)-X(\bd r,t)\right]d\bd{r}'.
\end{eqnarray*}
Correspondingly, the equation for the phase is modified as
\begin{equation}
\label{5b}
\partial_t\phi=\mbox{grad}_{\boldsymbol{A}}\phi\cdot 
(\bd{F}(\bd{A})+\bd{p})=\omega+\mbox{grad}_{\boldsymbol{A}}\phi\cdot \bd{p}.
\end{equation}
If the perturbation $\bd p$ is sufficiently weak, which we assume now,
the oscillator will keep staying on $\cal C$ in good
approximation. Then $\mbox{grad}_{\boldsymbol{A}}\phi$ in Eq.~\eqref{5b}
may safely be evaluated on $\cal C$, or
\begin{equation*}
\mbox{grad}_{\boldsymbol{A}}\phi\simeq (Z_{X}(\phi),Z_{Y}(\phi)),
\end{equation*}
where
\begin{equation*}
Z_{X}(\phi)=[\partial_{X}\phi(\bd{A})]_{\boldsymbol{A}=\boldsymbol{A}_{0}(\phi)},
\end{equation*}
and $Z_{Y}(\phi)$ is defined similarly.
At the same time, $p_X$ may be approximated with
\begin{equation*}
p_X\simeq K\int G(\bd{r}-\bd{r}')[X_{0}(\phi(\bd{r}',t))-X_{0}(\phi(\bd{r},t))]d\bd{r}'.
\end{equation*} 
Thus, the phase equation becomes
\begin{eqnarray*}
\partial_{t}\phi(\bd{r},t)&=&\omega+KZ_{X}(\phi(\bd{r},t))\\
&&\times\int G(\bd{r}-\bd{r}')[X_{0}(\phi(\bd{r}',t))-X_{0}(\phi(\bd{r},t))]d\bd{r}'.
\end{eqnarray*}
Since the small effect of the perturbation on $\partial_{t}\phi$ can be
time-averaged over one cycle of oscillation\cite{9}, the phase equation finally
takes the form
\begin{equation}
\label{7}
\partial_t\phi(\bd r,t)=\omega+K\int G(\bd{r}-\bd{r}')
\Gamma\big(\phi(\bd r,t)-\phi(\bd r',t)\big)d\bd{r}', 
\end{equation}
where
\begin{equation*}
\Gamma(\phi-\phi')=\frac{1}{2\pi}\int_{0}^{2\pi}
Z_X(\lambda+\phi)\left[X_{0}(\lambda+\phi')
-X_{0}(\lambda+\phi)
\right]d\lambda.
\end{equation*}
By using the above formula, $\Gamma(\phi)$ may be computed numerically if 
the forms of $f$ and $g$ are given explicitly.
For the present case of FitzHugh-Nagumo 
oscillators, numerically obtained $\Gamma(\phi)$  
is displayed in FIG.~\ref{fig3}.

The phase-coupling function $\Gamma(\phi)$, 
which is a $2\pi$-periodic function of $\phi$,
generally involves various harmonics, and
this is also true of the curve given in FIG.~\ref{fig3}. 
We still expect 
that the spiral dynamics of our concern does not depend
so heavily on the specific form of $\Gamma(\phi)$. 
Therefore, in order for further
mathematical analysis to be practicable, 
we will work with a simplest in-phase type 
coupling function, i.e., $\Gamma(\phi)=-\sin(\phi+\alpha)$ ($|\alpha|<\pi/2$). 
Thus, the phase equation takes the form
\begin{equation}
\label{8}
\partial_ t\phi(\bd{r},t)
=\omega-K\int G(\bd{r}-\bd{r}')\sin\big(\phi(\bd r,t)-\phi(\bd r',t)
+\alpha\big)d\bd{r}'
\end{equation}     
for which an in-depth mathematical analysis is possible as we see below.

Before proceeding to the analysis of Eq.~\eqref{8},
we remark that the above phase equation is  
also a correct reduced form of a
non-local version of the 
complex Ginzburg-Landau equation\cite{10}, the latter itself 
being a reduced form of 
our
three-component
reaction-diffusion model close to the Hopf bifurcation and comparably close
to the limit of vanishing coupling\cite{8}.
This fact gives a further support to our view that the application of
Eq.~\eqref{8} to our problem is reasonable.
 
We are still far from a full understanding of the solution to the
universal equation \eqref{8}, and
our concern below is its spiral wave solution
in two dimensions. Although the equation involves four parameters
$\omega$, $K$, $r_0$ and $\alpha$, the only relevant parameter is $\alpha$.
The reason is the following. Firstly, 
$r_{0}$, on which $G(r)$ depends (see Eq.~\eqref{9}), 
may be chosen to be the length unit, so that we may put
$r_{0}=1$. Similarly, the coupling strength $K$ may be fixed to 1 by suitably
choosing the time unit. The natural frequency $\omega$ can be eliminated 
by working with a suitable co-moving
frame of reference, i.e., via the transformation 
$\phi\rightarrow \phi+\omega t$.  In the following analysis, however,
the irrelevant parameter
$\omega$ is retained as
a nonzero constant, while we choose $\alpha=0.3$ and $r_0=K=1$. 

\begin{figure}
\includegraphics[width=8.6cm]{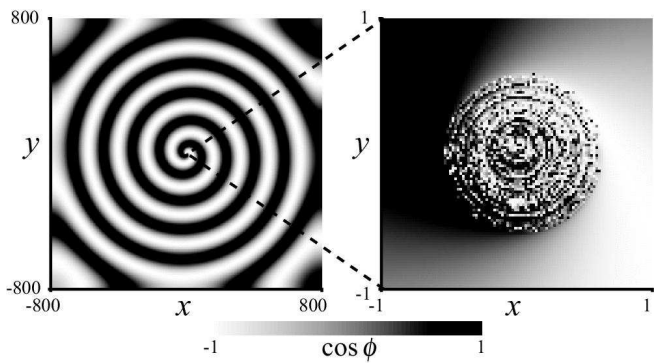}
\caption{\label{fig4} 
Spiral pattern (left) and its core structure (right) exhibited by 
non-locally coupled phase oscillators governed by
Eq.~\eqref{8}, where $\alpha=0.3$. }
\end{figure}

Numerical simulation of Eq.~\eqref{8} was carried out in a two-dimensional 
system. The numerical scheme adopted is the same as that explained in  the
previous section.
As expected, we see from FIG.~\ref{fig4} the appearance of 
rotating spiral waves 
with a disordered
group of oscillators near the core very similar to what we have seen
in the previous section.

For the arguments developed below,  
it is convenient to define a mean field 
$W(\bd{r},t)$ through
\begin{equation}
\label{10}
W(\bd{r},t)=\int G(\bd{r}-\bd{r}')\exp[i\phi(\bd{r}',t)]d\bd{r}'.
\end{equation}
The modulus $R$ and the phase
$\Theta$ of this complex quantity are defined by 
\begin{equation*}
W(\bd{r},t)=R(\bd{r},t)\exp[i\Theta(\bd{r},t)].
\end{equation*}
Since the definition \eqref{10} of the mean field 
involves a weighted spatial average over infinitely many
local oscillators, this quantity is expected to be smooth in space 
even these oscillators are behaving incoherently.
This property of $W$ is also clear from the differential form of 
Eq.~\eqref{10}, i.e., 
\begin{equation}
\label{10b}
0=-W+\nabla^2W+\exp(i\phi).
\end{equation}
The above equation implies a strong similarity of $W$ to $B$ governed by
Eq.~\eqref{3b}.
If the mean field pattern rotates steadily with frequency
$\Omega$,
then $R$ is time-independent and the relative mean field phase $\Theta_{0}$
defined by
\begin{equation*}
\Theta(\bd{r},t)=
\Omega t+\Theta_{0}(\bd{r},t)
\end{equation*}
is also time-independent.

In terms of $R(\bd{r},t)$ and $\Theta(\bd{r},t)$, 
Eq.~\eqref{8} may be expressed 
in the form
of a one-oscillator dynamics
\begin{equation*}
\partial_t\phi(\bd{r},t)=
\omega-R(\bd{r},t)\sin\big(\phi(\bd{r},t)+\alpha-\Theta(\bd{r},t)\big),
\end{equation*}
or if we introduce a relative phase variable $\psi(\bd{r},t)$ through
\begin{equation*}
\phi(\bd{r},t)=\Omega t+\psi(\bd{r},t),
\end{equation*}
we have 
\begin{equation}
\label{11}
\partial_t\psi(\bd{r},t)=
\omega-\Omega-R(\bd{r},t)\sin\big(\psi(\bd{r},t)+\alpha-\Theta_{0}(\bd{r},t)
\big).
\end{equation}
The definition of the mean field given by Eq.~\eqref{10} becomes
\begin{equation}
\label{12}
R(\bd{r},t)\exp[i\Theta_{0}(\bd{r},t)]=\int G(\bd{r}-\bd{r}')
\exp[i\psi(\bd{r}',t)]d\bd{r}'.
\end{equation}
Note that the set of Eqs.~\eqref{11} and \eqref{12} is still equivalent to the original phase
equation~\eqref{8}.

We now proceed to some {\em anatomy} 
of the anomalous core structure taking advantage
of the numerically observed fact that the mean field pattern has a well-defined
center of rotation (chosen to be $\bd r=0$) at which $W=0$.
One may thus imagine
a linear cross section $\cal S$ of the pattern passing through $\bd r=0$ and
study the radial profiles of various quantities emerging along $\cal S$.
Some results obtained in this way of analysis are summarized in FIG.~\ref{fig5} (a) to (c). 

An instantaneous radial profile of the mean field modulus $R$ is presented
in FIG.~\ref{fig5} (a). As expected, it has a 
vanishing value at the origin, and its temporal fluctuation
is also found negligibly small.

Figure~\ref{fig5} (b) shows an instantaneous distribution of the phases $\phi$
of the local oscillators lying on $\cal S$ (indicated by crosses
). The same panel also includes 
the pattern of the mean field phase $\Theta$ 
on the same 
cross
section (indicated by open circles). 
It is clear that there exists a well-defined critical
radius separating the domains of coherent and incoherent
oscillators from each other. We also confirmed (but not shown explicitly) 
that the profiles of the 
mean field phase and that of the phases of the coherent oscillators are
almost stationary except for a vertical drift with constant velocity $\Omega$.

Our interpretation of the results
given by FIGs.~\ref{fig5} (a) and (b) is that the entire system
now splits into two subdomains such that the oscillators in one domain
synchronize
completely with the periodic mean-field forcing, while those in the other
domain fail in
synchronization. Further evidence supporting this interpretation
is provided 
in FIG.~\ref{fig5} (c) where the distribution
of the mean frequency $\bar{\omega}$ 
(defined by a long-time average of $\partial_t\phi$)
of the local oscillators lying on $\cal S$ is shown.
This frequency pattern is clearly composed of two parts. Namely,
in the outer domain the oscillators have an identical frequency, while
in the inner domain the frequencies are distributed, the latter implying phase 
randomization consistent with the scattered dots appearing in 
FIG.~\ref{fig5} (b). 

In the next section, we develop a theory for determining the mean field pattern
together with its rotation frequency, and also the motion of the
individual oscillators driven by this mean field, in a self-consistent manner.

\section{THEORY}
The basic equations to work with are Eqs.~\eqref{11} and \eqref{12}.
Our theory starts with the assumption that the mean-field pattern is
steadily rotating, and therefore we drop the $t$-dependence from 
$R$ and $\Theta_{0}$
in these equations.
A complete solution to this system of equations can be obtained in the
following two steps. We first solve Eq.~\eqref{11} for each $\psi$ as 
a function of $R$ and $\Theta_{0}$, which is easy to do. 
Note that $R$ and $\Theta_{0}$ 
are the quantities yet to be
determined. Secondly, the entire set of these solutions is substituted into
Eq.~\eqref{12}. The right-hand side of Eq.~\eqref{12} thus becomes a functional of the
mean field. In this way, the mean field value at each spatial point is
expressed in terms of a functional of the mean field itself. 
Solution of this
functional self-consistency equation exists only for a special value of
the rotation frequency $\Omega$ of the mean field pattern. 
We will therefore be working with
a non-linear eigenvalue problem. The final solution of this functional 
equation could be found only numerically.

The above self-consistent way of finding a solution to a many-oscillator
problem resembles strongly Kuramoto's 1975 theory of synchronization
phase transition in a large population of globally coupled oscillators with
distributed natural frequencies\cite{5}. 
The main difference is that the oscillators
are now coupled non-locally rather than globally, and consequently
the mean field is generally
space-dependent leading to a functional
self-consistency equation rather than a simple transcendental equation.
Although the natural frequencies of the oscillators are 
identical in the present case,
the actual frequencies can be distributed due to the existence of
a spatial gradient of the mean field.
A simpler, one-dimensional version of the present type of theory 
based on a similar model of non-locally coupled phase oscillators was
reported earlier\cite{11}.

An important feature common to all such theories is that the 
one-oscillator 
equation which involves the mean field amplitude as a 
parameter
admits either a stationary solution or a drifting
solution. Which one to hold depends on the modulus of the mean-field.
The crucial point to the theory is how to deal with the drifting solutions,
because a simple substitution of this type of solutions
into the definition of the mean field apparently contradicts the
assumed stationarity of the mean field (in a suitable co-moving frame
of reference).  The seeming contradiction here can be resolved by using
the invariant
measure associated with the drift motion. We will now show explicitly
the steps leading to an exact solution to the problem.

As stated above,  
there are two possible cases regarding the solution of Eq.~\eqref{11}.
\begin{figure}
\includegraphics[width=8.6cm]{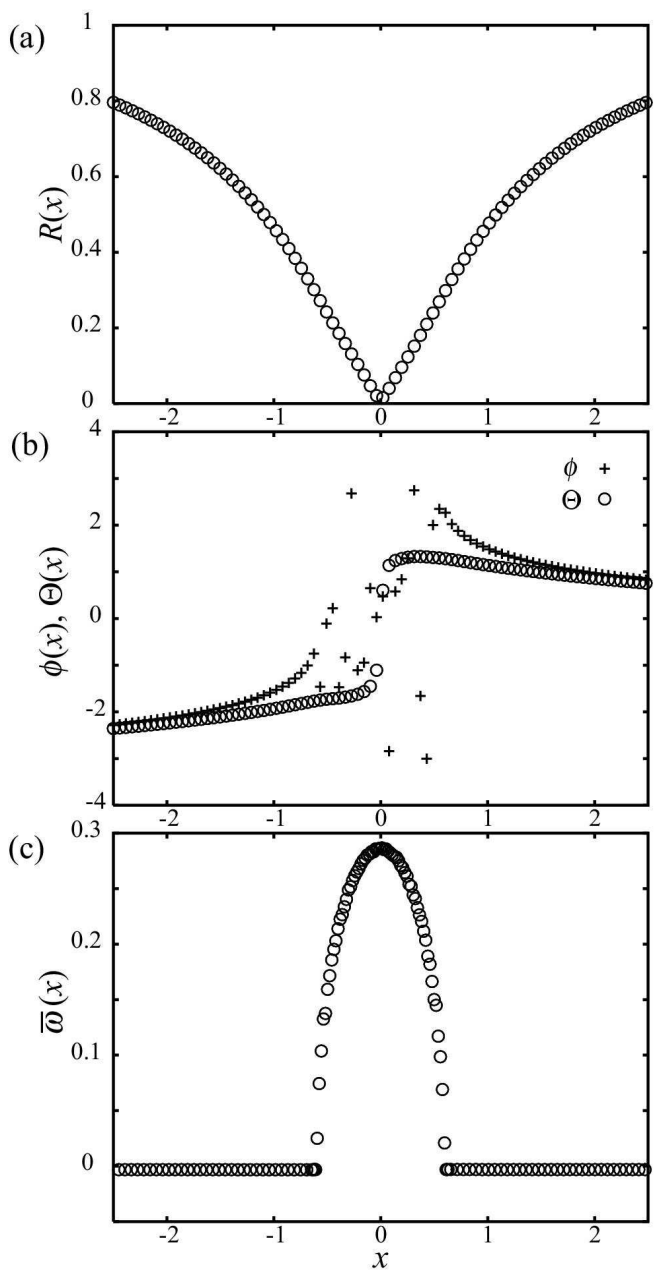}
\caption{\label{fig5}
Radial profiles of various quantities corresponding to
the spiral core of FIG.~\ref{fig4}.
(a) Instantaneous radial profile of the mean-field modulus $R$.
(b) Instantaneous radial profile of the phases $\phi$ of the local oscillators
(crosses) and that of the mean field phase $\Theta$
(open circles).
(c) Radial profile of the mean frequency $\bar{\omega}$ (defined by a 
long-time average of $\partial_t\phi$)
of the local oscillators.
} 
\end{figure}
They are: 
(Case I) $|\omega-\Omega|<R$, and (Case II) $|\omega-\Omega|>R$.
Correspondingly, the oscillators are divided into two groups.
In the first case, which corresponds to the group of coherent oscillators,
Eq.~\eqref{11} admits a pair of stable and unstable 
fixed points. The stable one, denoted by $\psi_0(\bd r)$, is given by
\begin{equation*}
\psi_0(\bd{r})=\mbox{Sin}^{-1}\Biggl( \frac{\omega-\Omega}{R(\bd{r})}\Biggr)
+\Theta_{0}(\bd{r})-\alpha.
\end{equation*}
The actual frequencies $\bar{\omega}$ of the oscillators in this group
are of course identical with $\Omega$.
We substitute the above solution for $\psi(\bd r)$ 
into Eq.~\eqref{12}, and restrict the integral
to the domain where the inequality $|\omega-\Omega|<R(\bd{r})$ is satisfied. 
In this
way, the contribution to the local mean field value coming 
from the coherent group of oscillators is obtained.

The second case corresponds to the group of incoherent oscillators, for which
Eq.~\eqref{11} admits a drifting solution. The actual frequencies 
$\bar{\omega}(\bd r)$ are now distributed and they are easily calculated as
\begin{equation*}
\begin{split}
\bar{\omega}&=\Omega+2\pi \Big[\int_{0}^{2\pi}
\Big(\frac{d\psi}{dt}\Big)^{-1}d\psi\Big]^{-1}\\
&=\Omega+(\omega-\Omega)\sqrt{1-\left(\frac{R}{\omega-\Omega}
\right)^{2}}.
\end{split}
\end{equation*}
The contribution to the local mean field value 
from this incoherent group of oscillators 
can be found in the following way.
Since $\psi$ is drifting,
the factor $\exp(i\psi)$ in the integrand in Eq.~\eqref{12} 
does not have a definite
value. We are thus led to the idea that this factor should rather be
replaced with its statistical average which can be
calculated by using the invariant measure, i.e., the probability density $p(\psi)$ 
associated with the drift 
motion. Noting that the probability density for the oscillator's phase 
to take on value
$\psi$ must be inversely proportional to the drift velocity given by the 
right-hand side of Eq.~\eqref{11}, we have
\begin{equation}
\label{14}
p(\psi)=C\left[\omega-\Omega-R
\sin\big(\psi+\alpha-\Theta_{0}\big)\right]^{-1},
\end{equation}
where $C$ is the normalization constant given by 
$C=(2\pi)^{-1}(\omega-\Omega)\sqrt{1-R^{2}/(\omega-\Omega)^{2}}$.

Putting together 
the above-stated two types of contributions to the mean field, 
we finally obtain a functional
self-consistency equation in the form
\begin{equation}
\label{13}
R(\bd r)e^{i\Theta_{0}(\bd r)}=
\int G(\bd{r}-\bd{r}')h(R(\bd r'),\Theta_{0}(\bd r'),\omega-\Omega)
d\bd{r}',
\end{equation}
where
\begin{equation*}
h(R,\Theta_0,\omega-\Omega)=\begin{cases}
	\exp[i\psi_0(R,\Theta_0,\omega-\Omega)]\\
	 \qquad\qquad\qquad\qquad\quad(\lvert\omega-\Omega\rvert<R), \\
	\int_{-\pi}^{\pi}p(\psi,R,\Theta_0,\omega-\Omega)
	\exp(i\psi)d\psi\\
	 \qquad\qquad\qquad\qquad\quad(\lvert\omega-\Omega\rvert>R),
                   \end{cases}
\end{equation*}
or more explicitly
\begin{equation*}
e^{i\psi_{0}}=e^{i(\Theta_0-\alpha)}
                    \left\{
                    \sqrt{1-\left(\frac{\omega-\Omega}{R}\right)^2}
                   +i\frac{\omega-\Omega}{R}\right\},
\end{equation*}
\begin{eqnarray*}
\int_{-\pi}^{\pi}p(\psi)e^{i\psi}d\psi
&=&i e^{i(\Theta_0-\alpha)}\left(\frac{\omega-\Omega}{R}\right)\\
	   &&\times
           \left\{1-\sqrt{1-\left(\frac{R}{\omega-\Omega}\right)^2}
           \right\}.
\end{eqnarray*}

Numerical solution of Eq.~\eqref{13} 
can be found iteratively. We did this in a 
finite domain defined by $x,y\in [0,40]$ with $G$ appropriate for the
free boundary conditions imposed on Eq.~\eqref{10b}.
Since a solution of Eq.~\eqref{13} would only be available 
for a special value of $\Omega-\omega$ 
which is still to be determined, its trial value was adjusted
in each
iteration step in such a way that a suitably defined distance between the
two mean-field patterns, one produced at the current step and the other
at the next step,
may be minimized.
In this way, by starting with a suitable initial mean field pattern
similar to the one obtained from numerical simulations, a rapid convergence
of the mean field pattern and the value of $\Omega$ was achieved. 

\begin{figure}
\includegraphics[width=8.6cm]{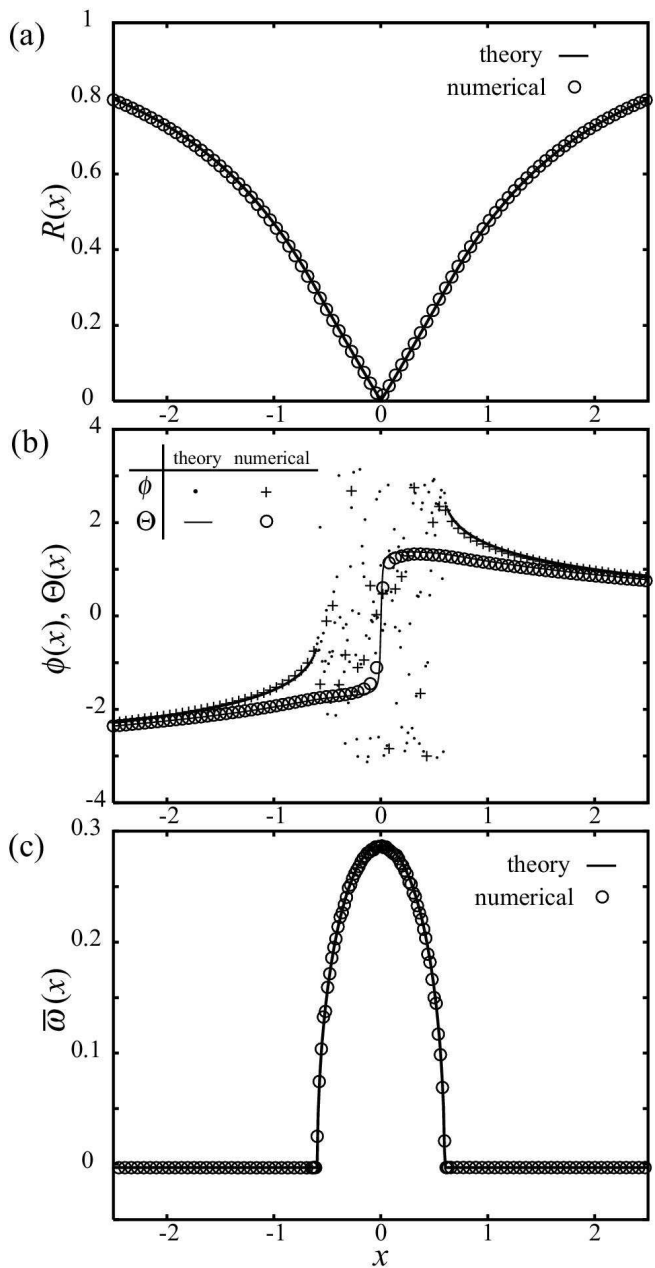}
\caption{\label{fig6} 
Comparison between the theory and numerical simulation.
Theoretical results are indicated with solid lines in (a) and (c),
and solid lines and scattered dots in (b). Numerical data, which
are the same as those given in FIG.~\ref{fig5}, are 
indicated with open circles and crosses. 
(a) Instantaneous radial profile of the mean-field modulus $R$. 
(b) Instantaneous radial profile of the mean-field phase $\Theta$, and
that of the phases $\phi$ of the local oscillators,
where the theoretically obtained scattered dots
are the random numbers chosen from the probability distribution 
given by
Eq.~\eqref{14}.
(c) Radial profile of the mean frequency $\bar{\omega}$ 
of the local oscillators.
}
\end{figure}

In FIG.~\ref{fig6} our theoretical results obtained 
in this way are compared with
the data given in FIG.~\ref{fig5}, i.e., the results from direct numerical
simulation of Eq.~\eqref{8}.  The agreement is so excellent
that our theory is expected to hold exactly in the continuum limit.

\section{SUMMARY AND CONCLUDING REMARKS}
Spontaneous generation of a local group of phase-randomized oscillators 
near the center
of a rotating spiral pattern
was confirmed through numerical simulations on non-locally coupled
oscillators. It was argued that smaller value of the coupling strength
favors the occurrence of the core anomaly.
The critical coupling strength $K_c$ associated with the onset 
of this
anomaly was estimated from a simple argument.
When $K$ is sufficiently small, by which the oscillation amplitude even 
near the
center of rotation takes almost a full value,  a group of incoherent
oscillators always exists. Still
the overall spiral pattern looks completely normal.
Guided by this fact observed numerically, 
we applied the phase reduction method for the purpose of
gaining a deeper understanding of the phenomenon.
The resulting phase oscillator model with non-local coupling 
was found to exhibit 
the same type of core anomaly.
Under the assumption 
that the pattern of a suitably defined mean field is steadily
rotating in spite of the existence of incoherence,
we derived a functional self-consistency equation to be satisfied by the
mean field. Its solution successfully reproduced various
results obtained from our direct 
numerical simulations carried out on this phase model.

Finally, we remark that the present study is confined to a particular
domain of parameter values where the mean field
dynamics is regular. Our preliminary study suggests that
under different conditions more complex collective dynamics occurs, which is 
characterized, e.g., by an elongation of the domain of incoherent oscillators
and its irregular motion\cite{1}. For the case of non-locally coupled 
FitzHugh-Nagumo
oscillators for small $K$, 
this occurs for larger $b$, i.e., when the symmetry of the local
oscillator dynamics is lowered, although a perfect symmetry ($b=0$, or the
van der Pol limit) is not necessary for the steady
rotation of the mean field. For larger $K$ (still below $K_c$), in contrast,
regular dynamics of the mean field and circular shape of the domain of 
incoherent oscillators seem to persist against relatively strong asymmetry of 
the oscillator dynamics. 
These results will be reported elsewhere.

\begin{acknowledgments}
The authors are grateful to H. Nakao for informative discussions.
\end{acknowledgments}


\bibliography{ref}

\end{document}